\documentclass[journal]{IEEEtran}

\usepackage{graphicx}
\graphicspath{{figures/}}
\usepackage{booktabs}
\usepackage{multirow}
\usepackage{placeins}
\usepackage{amsmath,amssymb}
\usepackage{cite}
\begin{document}

\title{Efficiency-Resolved Recovery Dynamics of an Free-Running InGaAs/InP Single Photon Avalanche Detector Operated in Gated Mode}

\author{Bruno~Santos~Souza, Vitor~Tavares, Filippo~Ghiglieno, Celso~Jorge~Villas-Boas,
and~Paulo~Henrique~Dias~Ferreira
\thanks{This work was supported in part by FAPESP under Grant 
2022/00209-6, and in part by the CNPq Quantum Dice project (Process No. 407665/2025-0), within the CNPq/MCTI Call No. 44/2024.}
\thanks{B. S. Souza, V. Tavares, C.J. Villas-Boas, and P. H. D. Ferreira are with the Departamento de F\'{i}sica, Universidade Federal de S\~{a}o Carlos, S\~{a}o Carlos, SP 13565-905, Brazil (corresponding author: Paulo Henrique Dias Ferreira, e-mail: paulohdf@df.ufscar.br).}
\thanks{F. Ghiglieno is with the Instituto de F\'{i}sica, Universidade de Bras\'{i}lia, Bras\'{i}lia, DF 70910-900, Brazil.}}

\maketitle

\begin{abstract}
Detection efficiency in gated-mode single-photon avalanche detectors recovers gradually after each gate edge, limiting attainable count rates, yet quantitative recovery models exist only for free-running operation. We extract the gated recovery time constant of a free-running InGaAs/InP detector at 1550~nm by sweeping the gate frequency from 100~kHz to 1~MHz and globally fitting count rates across four dead times, requiring neither a pulsed laser nor a time-to-digital converter. The recovery time decreases linearly with detection efficiency, from 301 to 161~ns between 10~\% and 25~\%, at 9.3~ns per percentage point. We further resolve a periodic count-rate modulation consistent with bias-dependent capacitive gate feedthrough.
\end{abstract}


\section{Introduction}

Single-photon avalanche diodes (SPADs) operating at telecom wavelengths constitute the workhorse detectors of practical quantum key distribution (QKD) systems. In these receivers, the achievable secret-key rate is ultimately limited by the detector count rate \cite{Huang2021JLT}, which is constrained not only by the imposed dead time but also by the recovery of detection efficiency following each detection event. As QKD systems evolve toward higher clock frequencies and smaller photon budgets, an accurate description of detector recovery becomes increasingly important for predicting count-rate saturation, optimizing operating conditions, and evaluating security-relevant detector effects \cite{Huang2021JLT,Weier2011,Makarov2006}.

The conventional description assumes an ideal detector that remains completely insensitive during a fixed dead time and instantaneously recovers its nominal quantum efficiency afterward \cite{Sarbazi2018,Zambon2022}. Although this step-function approximation successfully describes low-count-rate operation, it progressively overestimates the effective detection efficiency as the incident photon flux approaches the detector saturation regime.

Recently, Krause and Walenta demonstrated that this assumption is inadequate for free-running InGaAs/InP SPADs by introducing an exponential-recovery (ER) model, in which the detection efficiency recovers according to
\[
\eta(t)=\eta_0\left(1-e^{-t/\tau_r}\right),
\]
where $\tau_r$ is the recovery time constant determined by the recharge dynamics of the excess bias voltage through the detector capacitance and quenching circuit \cite{Krause2025}. Their model accurately reproduced inter-detection interval distributions over a photon-flux range approximately two orders of magnitude larger than the conventional step-function model and demonstrated that, for a detector operated at fixed efficiency, $\tau_r$ behaves as an intrinsic parameter independent of the incident photon rate \cite{Krause2025}. Exponential recovery is therefore an essential ingredient in accurate count-rate models for free-running SPADs. Whether and how it manifests under gated operation \cite{Mahmoudi2021,Altilia2026,Gebremicael2021,Fan2023,Huang2021JLT,Losev2022PJ,Losev2022JQE,Itzler2012}, where the efficiency reset is imposed by the gate edge rather than by the detection event itself, has not been established.

Practical QKD systems, however, commonly employ gated InGaAs/InP SPADs, in which the detector is periodically biased above breakdown only during narrow time windows synchronized with expected photon arrival times. In this operating mode, the detector undergoes repeated externally imposed bias excursions, making the recovery process fundamentally different from free-running operation. Recovery is no longer continuously sampled after each detection event, but only at discrete gate openings determined by the system clock. Consequently, the detector response depends on a complex interplay between dead time, gate width, gate repetition frequency, and excess bias voltage. Previous studies have investigated gated-SPAD timing behavior, avalanche build-up, dark counts, afterpulsing, and effective breakdown voltage under gated operation \cite{Mahmoudi2021,Altilia2026,Gebremicael2021,Fan2023,Huang2021JLT}. Nevertheless, none of these studies directly characterizes the recovery dynamics itself, nor do they address whether the recovery time remains invariant as the detector operating efficiency changes.

This distinction is particularly critical because, unlike photon flux, detection efficiency is directly tuned via the excess bias voltage ($V_{\text{ex}}$). Since both avalanche triggering probability and junction recharge originate from the same bias-dependent physical mechanisms, it is not obvious that the recovery time constant should remain unchanged when detector efficiency is varied. To the best of our knowledge, this dependence has not been experimentally investigated for gated InGaAs/InP SPADs.

This work addresses this gap by introducing a simple frequency-sweep method under continuous-wave (CW) illumination that directly extracts the recovery time constant of gated SPADs without requiring a high-speed pulsed laser or a time-to-digital converter (TDC). By measuring the detector count rate as a function of gate frequency across multiple detection efficiencies on a commercial IDQube NIR InGaAs/InP detector, we extract the recovery dynamics through global fits of the measured count rate response. 

Our measurements reveal a key physical result: the recovery time constant decreases systematically as detection efficiency increases. While Krause and Walenta established that recovery time is independent of photon flux at a fixed efficiency in free-running mode \cite{Krause2025}, our results show that, under gated operation, the recovery time constant depends on the excess bias voltage set by the detector's  efficiency selection, decreasing as efficiency increases, and is common across dead time settings, indicating that it is governed by the bias point rather than by the externally selected dead time. Additionally, we observe a subtle periodic modulation superimposed on the count-rate curves, whose amplitude decreases with increasing detection efficiency, consistent with capacitive gate feedthrough modulated by bias-dependent junction capacitance. Together, these findings provide new insight into the internal physics of gated InGaAs/InP SPADs and provide a lightweight characterization framework directly applicable to practical QKD receiver optimization.

\section{Recovery model for gated SPADs}
\label{sec:model}

The analytical framework developed in this work relies on three physical premises:
\begin{enumerate}
    \item Following detector re-activation or gate turn-on, the detection efficiency does not recover instantaneously, but approaches its steady-state value continuously via excess-bias recharge.
    \item The mean detection probability per gate window is determined by the time-integrated recovery profile over the gate duration.
    \item Upon avalanche detection, the quenching electronics enforce an unyielding dead time, after which the SPAD re-arms only at the discrete onset of the subsequent gate pulse.
\end{enumerate}
Together, these conditions yield a self-contained count-rate model parameterized by a single intrinsic recovery time constant, $\tau_{\mathrm{rec}}$.

\subsection{Exponential recovery dynamics}

\begin{figure}[htbp]
\centering
\includegraphics[width=\columnwidth]{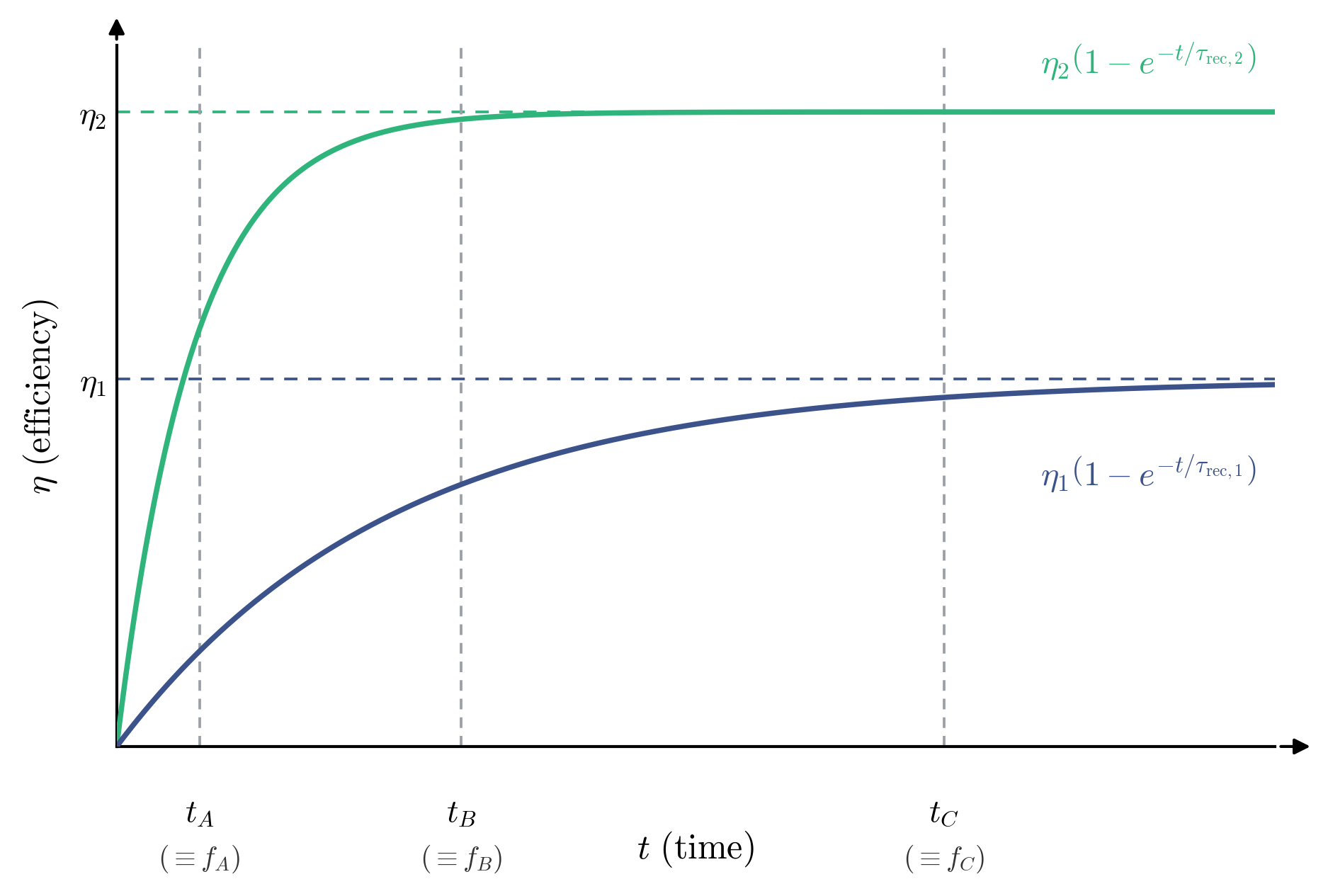}
\caption{Schematic exponential recovery of the local detection efficiency $\eta(t)$ toward its asymptotic value $\eta_0$ [Eq.~(\ref{eq:eta})], shown for two representative operating points, $\eta_1 < \eta_2$, with $\tau_{\mathrm{rec},2} < \tau_{\mathrm{rec},1}$. The curves are drawn schematically, not fitted to data, to isolate the qualitative effect. Vertical lines mark three gate durations $t_A < t_B < t_C$, equivalently three gate frequencies $f_A > f_B > f_C$ through $t = W = D/f$: at short gate duration ($t_A$) neither curve has saturated, while at long gate duration ($t_C$) both approach their respective plateaus.}
\label{fig:eta_schematic}
\end{figure}

When the imposed inhibition interval expires, and/or when a new gating pulse is
applied, the excess bias voltage $V_{\mathrm{ex}}(t)$ across the SPAD junction
does not instantly step to its nominal operating level. Driven by the passive or active recharge dynamics of the quenching loop and junction capacitance $C_j$ \cite{Cova1996,Inoue2020,Krause2025}, the local detection efficiency $\eta(t)$ recovers exponentially toward its asymptotic maximum $\eta_0$:
\begin{equation}
\frac{\eta(t)}{\eta_0} = 1 - \exp\left(-\frac{t}{\tau_{\mathrm{rec}}}\right), \qquad 0 \le t \le W,
\label{eq:eta}
\end{equation}
where $W = D/f$ represents the gate duration at repetition frequency $f$ and fixed electronic duty cycle $D$, while $\tau_{\mathrm{rec}}$ is the characteristic recovery time constant. Figure~\ref{fig:eta_schematic} illustrates this recovery for two representative efficiency settings, $\eta_1 < \eta_2$, with $\tau_{\mathrm{rec},2} < \tau_{\mathrm{rec},1}$, anticipating the trend established quantitatively in Sec.~\ref{sec:tau_results}: detectors biased for higher detection efficiency recover faster. Because the gate remains open for only a finite duration $W = D/f$, the detector never has unlimited time to complete the exponential approach to $\eta_0$; each gate instead samples a truncated segment of $\eta(t)$ whose extent is set by the gate frequency. At high frequency (short $W$, e.g.\ $t_A$ in Fig.~\ref{fig:eta_schematic}), both curves are still rising, so the efficiency realized within the gate falls well below $\eta_0$, more severely so for the slower-recovering setting. At low frequency (long $W$, e.g.\ $t_C$), the gate stays open long enough for even the slower curve to saturate, and the two efficiencies converge toward their respective plateaus. This frequency-dependent truncation of $\eta(t)$, and its dependence on the recovery time constant, is what the gate-frequency sweep exploits to extract $\tau_{\mathrm{rec}}$ from count-rate data alone, formalized below as the recovery integral $I(f)$ [Eq.~(\ref{eq:If})].

Unlike free-running SPADs, where bias recovery begins after a user-defined dead time $\tau_{\mathrm{dt}}$ after an avalanche quenching event, the detector used here is operated in gated mode, in which an external gate signal periodically modulates the detector bias. Although the quenching electronics enforce a dead time $\tau_{\mathrm{dt}}$ after each detection, full recovery can only resume on the subsequent gate edge. Consequently, a detection occurring late within a gate pulse leaves the detector insensitive well past $\tau_{\mathrm{dt}}$ until the next gating pulse re-enables it, a mechanism analyzed in detail in Sec.~\ref{sec:full}.

Under this gated regime, $\tau_{\mathrm{rec}}$ directly characterizes the driven transient response of the junction architecture. Since the baseline detection efficiency $\eta_0$ acts as a global scale factor, it is absorbed into the effective photon arrival rate:
\begin{equation}
R_p \equiv \eta_0 R_{\mathrm{ph}},
\end{equation}
where $R_{\mathrm{ph}}$ is the absolute incident photon flux. This is the same construction as the a priori detection rate $R^* \equiv \eta_0 R_i$ introduced by Krause and Walenta for free-running SPADs \cite{Krause2025}, denoting the rate that would be registered in the absence of dead time and with instantaneous efficiency recovery; here $R_p$ plays the analogous role of an idealized reference rate, subsequently reshaped by the finite recovery integral $I(f)$ of Eq. (\ref{eq:If}) rather than by the simple dead-time relation of their Eq.~(1). The predicted count rate will therefore depend on two independent factors: an overall amplitude set by $R_{\mathrm{p}}$, and a shape factor set by how much of each gate window sees a fully recovered detector, which we quantify next.

\subsection{Effective gate window response}

Because a gated SPAD integrates carrier generation over the active gate window $W$, the effective detection capability is proportional to the time integral of the normalized recovery profile in Eq.~(\ref{eq:eta}):
\begin{equation}
I(f) = \int_0^W \frac{\eta(t)}{\eta_0} \, dt = W - \tau_{\mathrm{rec}} \left(1 - e^{-W/\tau_{\mathrm{rec}}}\right).
\label{eq:If}
\end{equation}
Physically, $I(f)$ represents an \emph{effective gate width}: the equivalent duration of a hypothetically ideal, fully recovered gate ($\eta = \eta_0$) yielding the same integrated detection probability. As $f$ increases, $W = D/f$ shrinks relative to $\tau_{\mathrm{rec}}$, causing $I(f)$ to decline nonlinearly and driving the count-rate saturation behavior evaluated in this study. (see Supplementary Material Sec.~S1 for the normalization removing $\eta_0$ from this integral).

\subsection{Models}

\subsubsection{Model B: Baseline count-rate model}

Under low mean photon flux per gate ($R_p I(f) \ll 1$), the probability of a detection occurring during a single gate is approximated by $p(f) = R_p I(f)$. Assuming an unconstrained, non-paralyzable dead time $\tau_{\mathrm{dt}}$ \cite{Cova1996,Sarbazi2018}, the baseline count-rate response $C_{\mathrm{B}}(f)$ reads:
\begin{equation}
C_{\mathrm{B}}(f) = \frac{f \, R_p \, I(f)}{1 + R_p \, I(f) \, \tau_{\mathrm{dt}} \, f}.
\label{eq:CB}
\end{equation}
Equation~(\ref{eq:CB}) correctly captures both fundamental asymptotic limits: low-flux linear scaling ($C_{\mathrm{B}} \to f R_p I(f)$) and high-flux dead-time saturation ($C_{\mathrm{B}} \to 1/\tau_{\mathrm{dt}}$).

\subsubsection{Model F: Full gated-detector model with discrete re-arming and gate ripple}
\label{sec:full}

Equation~(\ref{eq:CB}) assumes a continuous recovery process free of residual electronic coupling. Two physical effects violate this idealization: the discreteness of the gate edges can leave a small fraction of the recovery time unresolved between successive gates, and capacitive feedthrough from the gate signal imprints a periodic ripple on the measured count rate. We incorporate both into a full model, $C_{\mathrm{F}}(f)$.

First, gated re-arming is inherently discrete: following a detection occurring at time $t_{\mathrm{click}}$ within a gate, the detector cannot re-arm until the first gate opening following the expiration of $\tau_{\mathrm{dt}}$. Evaluating the mean click position weighted by the time-dependent efficiency $\eta(t)$ yields the closed-form expression (see Supplementary Material Sec.~S1 for the derivation of the underlying click-time probability density):
\begin{equation}
\langle t_{\mathrm{click}} \rangle = \frac{\int_0^W t \, \eta(t) \, dt}{I(f)} = \frac{\frac{W^2}{2} - \tau_{\mathrm{rec}}^2 \left[ 1 - \left(1 + \frac{W}{\tau_{\mathrm{rec}}}\right) e^{-W/\tau_{\mathrm{rec}}} \right]}{W - \tau_{\mathrm{rec}} \left(1 - e^{-W/\tau_{\mathrm{rec}}}\right)}.
\label{eq:tclick_analytic}
\end{equation}
Because the ceiling in the quantization below depends on the exact click
time within the gate, an exact treatment would require averaging the
quantized dead time over the full click-time distribution
$p_{\mathrm{click}}(t)$ (Supplementary Material Sec.~S1) for every gate.
We instead evaluate the quantization at the mean click position
$\langle t_{\mathrm{click}} \rangle$, a mean-field approximation that
becomes exact whenever the ceiling is constant across the support of
$p_{\mathrm{click}}(t)$, as is the case for the present measurement grid
(Sec.~\ref{sec:losttime_results}). The discrete effective dead time
$\tau_{\mathrm{eff}}(f)$ quantized to integer gate clock periods $T = 1/f$
is then written as:
\begin{equation}
\tau_{\mathrm{eff}}(f) = \left( \left\lceil \frac{\tau_{\mathrm{dt}} + \langle t_{\mathrm{click}} \rangle}{T} \right\rceil - 1 \right) T,
\label{eq:taueff}
\end{equation}
where $\lceil \cdot \rceil$ represents the ceiling function. This quantization
is not merely a modelling assumption: the manufacturer specifies that a gate
arriving while the detector is still inhibited is blanked, the device
resuming only at the next available gate \cite{IDQubeManual}. Physically,
$n-1$ counts the number of entire gate cycles for which the detector remains
blind after the clicking gate itself, and $\tau_{\mathrm{eff}}(f)$ expresses
this count as a time via $T$; it is not the continuous blind duration
measured from $t_{\mathrm{click}}$ to re-arming, which would instead be
$nT - t_{\mathrm{click}}$. This distinction matters only for bookkeeping: it
is the count of lost gate cycles, not the continuous duration, that enters
the non-paralyzable rate equation in the same role $\tau_{\mathrm{dt}}$ plays
in Eq.~(\ref{eq:CB}).

Second, fast gate voltage transients capacitively couple into the avalanche readout channel via the voltage-dependent junction capacitance $C_j(V_{\mathrm{ex}})$, inducing a weak, periodic modulation on the threshold discriminator response as a function of clock frequency \cite{Gebremicael2021,Fan2023}. We model this inter-gate feedthrough phenomenologically by modulating the single-gate detection parameter:
\begin{equation}
p(f) = R_p \, I(f) \left[ 1 + a \sin\left(\frac{2\pi f}{f_0} + \phi \right) \right],
\label{eq:pf}
\end{equation}
yielding the complete count-rate model $C_{\mathrm{F}}(f)$:
\begin{equation}
C_{\mathrm{F}}(f) = \frac{f \, p(f)}{1 + p(f) \, \tau_{\mathrm{eff}}(f) \, f}.
\label{eq:CF}
\end{equation}

In global multi-dataset fits across various dead-time settings $\tau_{\mathrm{dt}}$, the physical recovery time $\tau_{\mathrm{rec}}$ and ripple parameters $(a, f_0, \phi)$ are treated as shared intrinsic constants for a given operating efficiency $\eta_0$, whereas $R_p$ acts as a dataset-specific nuisance parameter.

\subsection{Physical origin of recovery-bias dependence}
\label{sec:ripple}

Both the recovery time constant $\tau_{\mathrm{rec}}$ and the feedthrough ripple parameters $(a, f_0, \phi)$ stem from the electrical dynamics of the gated SPAD junction. The recovery rate $\tau_{\mathrm{rec}}$ is governed by the $R_q C_j$ charging dynamic of the depleted junction through the quenching/bias network. Because increasing the nominal detection efficiency $\eta_0$ requires a higher excess bias voltage $V_{\mathrm{ex}}$, it alters both the instantaneous junction capacitance $C_j(V_{\mathrm{ex}})$ and the avalanche breakdown probability profile across the multiplication layer \cite{Jiang2007,Losev2022PJ}. Consequently, $\tau_{\mathrm{rec}}$ is intrinsically efficiency-dependent in gated operation—a core feature demonstrated experimentally in this work.

\section{Experimental Methods}
\label{sec:methods}

\subsection{Setup and acquisition protocol}

Measurements were performed on a commercial InGaAs/InP single-photon avalanche detector (ID Quantique, IDQube-NIR-FR-MMF-LN) operated in gated mode via its external Gate IN input, under continuous-wave illumination at 1550~nm. The detection efficiency is determined by the excess bias voltage $V_{\mathrm{ex}}$; we report four nominal values, $\eta \in \{10, 15, 20, 25\}\,\%$, set through the manufacturer's interface and not independently calibrated, so the linear $\tau_{\mathrm{rec}}(\eta)$ trend reported in Sec.~\ref{sec:results} assumes this scale to be linear. At each efficiency, the electronically selectable dead time was cycled through $\tau_{\mathrm{dt}} \in \{10, 20, 40, 80\}\,\mu$s with all other settings unchanged, since dead time is a quenching-circuit parameter independent of bias.

A terminological note: in the standard SPAD literature, total dead time sums the avalanche sensing, quenching, hold-off, and reset times, of which only the hold-off is user-programmable \cite{Cova1996,Ceccarelli2021}. The instrument used here exposes a single programmable parameter, which the manufacturer labels dead time and defines as the interval during which the SPAD is held below breakdown after each detection \cite{IDQubeManual}; this corresponds to the hold-off above. We write it as $\tau_{\mathrm{dt}}$, following the manufacturer's label, so that the values quoted here match the instrument settings directly.

The gate was driven by an external pulse generator at frequency $f$ with fixed duty cycle $D = 0.5$, giving the gate window $W = D/f$ of Sec.~\ref{sec:model}; sweeping $f$ sweeps the fraction of each gate spent in a partially recovered state, since the recovery of Eq.~(\ref{eq:eta}) is re-initiated at every gate edge. A continuous-wave 1550~nm source, attenuated through a fiber-coupled attenuator chain, provided the illumination, and count rates were logged with the ID Quantique IDQube acquisition software.

For each efficiency, $f$ was swept from 100 to 1000~kHz in 100~kHz steps, giving a gate window $W = D/f$ ranging from 500~ns to 5~$\mu$s. Combined with the four dead-time settings, this yields $N = 40$ measurement conditions per efficiency block, 160 in total. Each condition was recorded as the mean of $N_t = 69$ repeated acquisitions collected over approximately one minute, together with the corresponding sample standard deviation. Count rates ranged from 835 to 5537~cps, reaching up to 31~\% of the dead-time-limited ceiling $1/\tau_{\mathrm{dt}}$ at the longest dead time and highest efficiency; because Eq.~(\ref{eq:CB}) is the exact non-paralyzable expression rather than a low-flux truncation, this does not compromise the fit, though it reduces the sensitivity of $C(f)$ to $R_p$ near saturation, contributing to the mild degradation of $\chi^2_\nu$ with efficiency noted in Sec.~\ref{sec:discussion}.

\subsection{Global fitting strategy}

Parameters were estimated by weighted nonlinear least squares, equivalent to Gaussian maximum likelihood, jointly fitting all four dead-time datasets within an efficiency block and minimizing
\begin{equation}
\chi^2 = \sum_{j=1}^{N} \frac{\left[C_{\mathrm{obs},j} - C_{\mathrm{pred},j}(\boldsymbol{\theta})\right]^2}{\sigma_j^2},
\label{eq:chi2}
\end{equation}
where $\sigma_j$ is the standard error of the mean over the $N_t$ repetitions at point $j$, and $C_{\mathrm{pred},j}$ follows from Eq.~(\ref{eq:CB}) or Eq.~(\ref{eq:CF}). This treatment assumes the $N_t$ repetitions are statistically independent; as a consistency check, the ratio of the measured standard deviation to the Poisson expectation lies between 0.90 and 0.96 across the four efficiencies, below unity as expected for a dead-time-regularized counting process \cite{Zambon2022,Sarbazi2018}.

Within each efficiency block the four dead-time datasets are fitted simultaneously with a shared $\tau_{\mathrm{rec}}$, since recovery is a property of the junction and its quenching network rather than of the dead-time logic downstream; sharing it gives the parameter the leverage of all four curves at once. The effective photon rate $R_p$ is left free per dead time, as a nuisance parameter absorbing small drifts in optical coupling or rate calibration between settings, so that such drifts cannot bias $\tau_{\mathrm{rec}}$. The method requires only that illumination remain stable within each frequency sweep, as ensured by the acquisition protocol above. In the Full Model the ripple parameters $\{a, f_0, \varphi\}$ are likewise shared within a block, since they are governed by the same excess-bias-dependent mechanism (Sec.~\ref{sec:ripple}), but are free between blocks because $V_{\mathrm{ex}}$ differs. This gives $k_{\mathrm{S}} = 5$ and $k_{\mathrm{F}} = 8$ free parameters per efficiency block. Optimization used a trust-region reflective algorithm \cite{Virtanen2020}; because $C_{\mathrm{F}}(f)$ is periodic in $f_0$ and $\varphi$ its cost surface is multimodal, so local refinement was preceded by a coarse scan over $(f_0, \varphi)$ and initialized from the lowest-cost grid point. Fits were verified to be stable against $\pm 10\,\%$ perturbations of the starting parameters.

\subsection{Model comparison and residual diagnostics}
\label{sec:aicbic}

The Baseline and Full models were ranked by the Akaike and Bayesian information criteria,
\begin{equation}
\mathrm{AIC} = \chi^2 + 2k, \qquad \mathrm{BIC} = \chi^2 + k \ln N,
\label{eq:aicbic}
\end{equation}
which penalize the three additional parameters of the Full Model and so
guard against rewarding overfitting. We define $\Delta\mathrm{AIC} \equiv \mathrm{AIC}_{\mathrm B} - \mathrm{AIC}_{\mathrm F}$ and $\Delta\mathrm{BIC} \equiv \mathrm{BIC}_{\mathrm B} - \mathrm{BIC}_{\mathrm F}$, so that a positive value favors the Full Model; following convention, $\Delta\mathrm{AIC} > 10$ is treated as decisive \cite{BurnhamAnderson2004}. The reduced chi-square $\chi^2_\nu = \chi^2/(N-k)$ is reported separately as a measure of absolute fit quality.

Normalized residuals were inspected against gate frequency for systematic structure. Because the principal difference between the two models is a periodic term, we additionally computed the Lomb-Scargle periodogram \cite{VanderPlas2018} of the Baseline residuals: agreement between its dominant peak and the independently fitted $f_0$ provides evidence for the ripple that does not depend on the sinusoidal parameterization itself.

\subsection{Commensurability of the measurement grid}
\label{sec:losttime}

On the frequency and dead-time grid used here, $\tau_{\mathrm{dt}}/T$ is an integer at every one of the 160 measurement conditions, so $\tau_{\mathrm{eff}} = \tau_{\mathrm{dt}}$ identically; the consequences for model comparison are discussed quantitatively in Sec.~\ref{sec:losttime_results}.



\section{Results and Discussion}
\label{sec:results}
\label{sec:discussion}

The dataset comprises 160 $(\eta, \tau_{\mathrm{dt}}, f)$ conditions, 40 per detection efficiency (Sec.~\ref{sec:methods}). We present first the recovery dynamics, which are the principal result of this work, then the periodic modulation of the count rate, and finally the evidence that validates the model used to extract both. Each result is interpreted where it appears rather than in a separate section, and the scope of what the present grid can and cannot establish is stated at the end.

\subsection{Recovery dynamics as a function of detection efficiency}
\label{sec:tau_results}

\begin{figure}[htbp]
\centering
\includegraphics[width=\columnwidth]{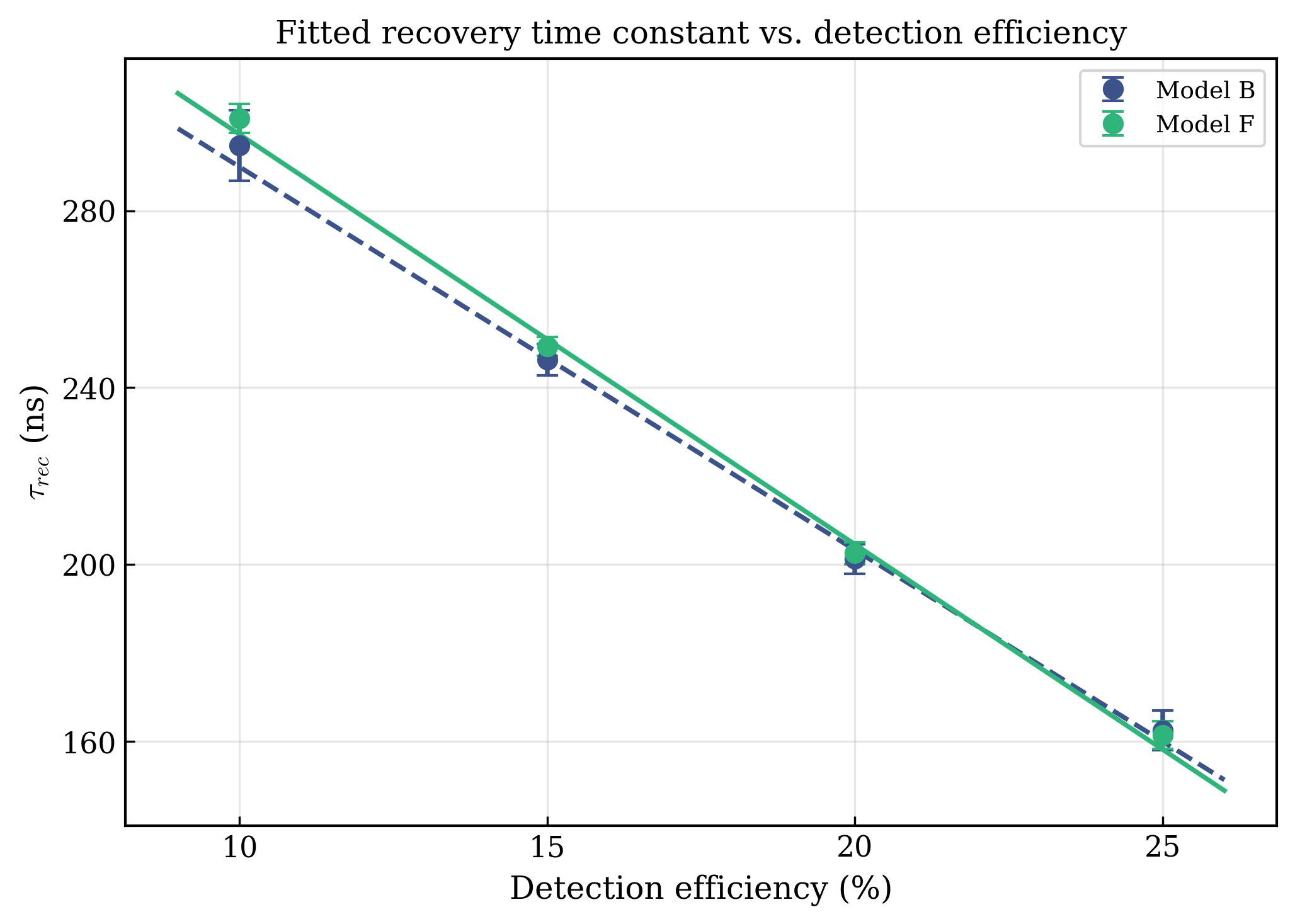}
\caption{Recovery time constant $\tau_{\mathrm{rec}}$ versus detection efficiency for the Baseline (B) and Full (F) models, $C_{\mathrm{B}}(f)$ and $C_{\mathrm{F}}(f)$, with the weighted linear fit of Eq.~(\ref{eq:taulin}).}
\label{fig:tau_vs_eff}
\end{figure}

Figure~\ref{fig:tau_vs_eff} summarizes the central finding: the gated recovery time constant shortens as the detection efficiency, and therefore the excess bias $V_{\mathrm{ex}}$, is raised.

\begin{table*}[!t]
  \centering
  \caption{Global fit results for the gated frequency-sweep characterization
  of the InGaAs/InP SPAD. Model B uses the nominal dead time only
  ($k=5$ free parameters); Model F adds the lost-time quantization (no extra
  free parameters) and the ripple term ($k=8$, three additional free
  parameters). Uncertainties on $\tau_{rec}$ are inflated by
  $\sqrt{\chi^2_\nu}$.}
  \label{tab:global_fits}
  \begin{tabular}{c l c c c c c c}
    \toprule
    $\eta$ (\%) & Model & $\tau_{rec}$ (ns) & $R^2$ & $\chi^2$ & $\chi^2_\nu$ & AIC & BIC \\
    \midrule
    \multirow{2}{*}{10} & B & 294.7 $\pm$ 8.0 & 0.9817 & 2482.2 & 70.92 & 2492.2 & 2500.6 \\
     & F & 300.9 $\pm$ 3.3 & 0.9988 & 170.8 & 5.34 & 186.8 & 200.4 \\
    \addlinespace
    \multirow{2}{*}{15} & B & 246.3 $\pm$ 3.5 & 0.9940 & 833.4 & 23.81 & 843.4 & 851.8 \\
     & F & 249.3 $\pm$ 2.1 & 0.9988 & 195.3 & 6.10 & 211.3 & 224.8 \\
    \addlinespace
    \multirow{2}{*}{20} & B & 201.3 $\pm$ 3.4 & 0.9930 & 1204.9 & 34.43 & 1214.9 & 1223.4 \\
     & F & 202.5 $\pm$ 2.5 & 0.9979 & 369.3 & 11.54 & 385.3 & 398.8 \\
    \addlinespace
    \multirow{2}{*}{25} & B & 162.5 $\pm$ 4.5 & 0.9889 & 3516.6 & 100.47 & 3526.6 & 3535.0 \\
     & F & 161.4 $\pm$ 3.2 & 0.9980 & 691.5 & 21.61 & 707.5 & 721.0 \\
    \bottomrule
  \end{tabular}
\end{table*}

The dependence is close to linear across the full range studied. A weighted linear fit to the Full Model values gives
\begin{equation}
\tau_{\mathrm{rec}}(\eta) = (390.1 - 9.28\,\eta)~\mathrm{ns},
\label{eq:taulin}
\end{equation}
where $\eta$ is the detection efficiency in percentage points and the slope, $-9.28 \pm 0.28$~ns per percentage point (weighted $R^2 = 0.997$), quantifies the acceleration of recovery with increasing bias. The Baseline Model returns statistically indistinguishable coefficients (Table~\ref{tab:global_fits}). All quoted uncertainties are inflated by $\sqrt{\chi^2_\nu}$, following standard practice when the reduced chi-square exceeds unity.

This trend follows the recovery mechanisms motivated in Sec.~\ref{sec:model}. A larger excess bias strengthens the electric field across the junction, accelerating both the escape of retrapped carriers and the capacitive recharge of the depletion region \cite{You2012, Inoue2020,Jiang2007}, so $\tau_{\mathrm{rec}}$ shortens as $V_{\mathrm{ex}}$, and with it $\eta$, is raised. The magnitude of the effect is operationally significant rather than marginal: a detector configured for $\eta = 25$ \% re-arms within each gate nearly twice as fast as the same detector at $\eta = 10$ \%, a $46$ \% reduction in recovery time over the range studied. The choice of operating point therefore alters not only the probability that an incident photon is registered but also the photon rate at which recovery begins to limit throughput. Any treatment of gated detection efficiency as a single frequency-independent number, which is the standard assumption in gated count-rate models, discards this dependence entirely.


We attribute this systematic offset, present at every efficiency tested, not to a discrepancy to be reconciled but to the two time constants being anchored to different triggering events. $\tau_r$, the free-running recovery constant reported by Krause and Walenta \cite{Krause2025} for this same detector model and discussed quantitatively in Sec.~\ref{sec:scope}, is anchored directly to the detection event: recovery begins the instant the avalanche is quenched, with no external reference. $\tau_{\mathrm{rec}}$, by contrast, is anchored to the gate edge, since Eq.~(\ref{eq:eta}) resets $\eta(t)$ to zero at the start of each gate regardless of when within the preceding dead interval the last detection occurred. Gated recovery must therefore additionally accommodate the interval between a detection and the next gate edge, an interval with no counterpart in free-running operation, which accounts for why $\tau_{\mathrm{rec}} > \tau_r$ is the systematic pattern observed here. The two characterizations are complementary rather than competing: $\tau_{\mathrm{rec}}$ governs count-rate performance in any system that operates the detector in gated mode, which includes essentially all deployed QKD receivers, while $\tau_r$ characterizes the intrinsic junction recovery independent of external gating and is the quantity relevant to ungated applications.

The extracted recovery time is furthermore insensitive to the modulation correction introduced below. The two models agree on $\tau_{\mathrm{rec}}$ at every efficiency, with a maximum discrepancy of $0.73\sigma$ and a largest absolute difference of $6.1$~ns at $\eta = 10\,\%$, well within the combined uncertainty of $8.6$~ns. The central result thus rests only on the recovery integral of Eq.~(\ref{eq:If}) together with the non-paralyzable dead-time relation, and not on how, or whether, the ripple is parameterized.

\subsection{Periodic modulation of the count rate}
\label{sec:ripple_results}

\begin{figure}[htbp]
\centering
\includegraphics[width=\columnwidth]{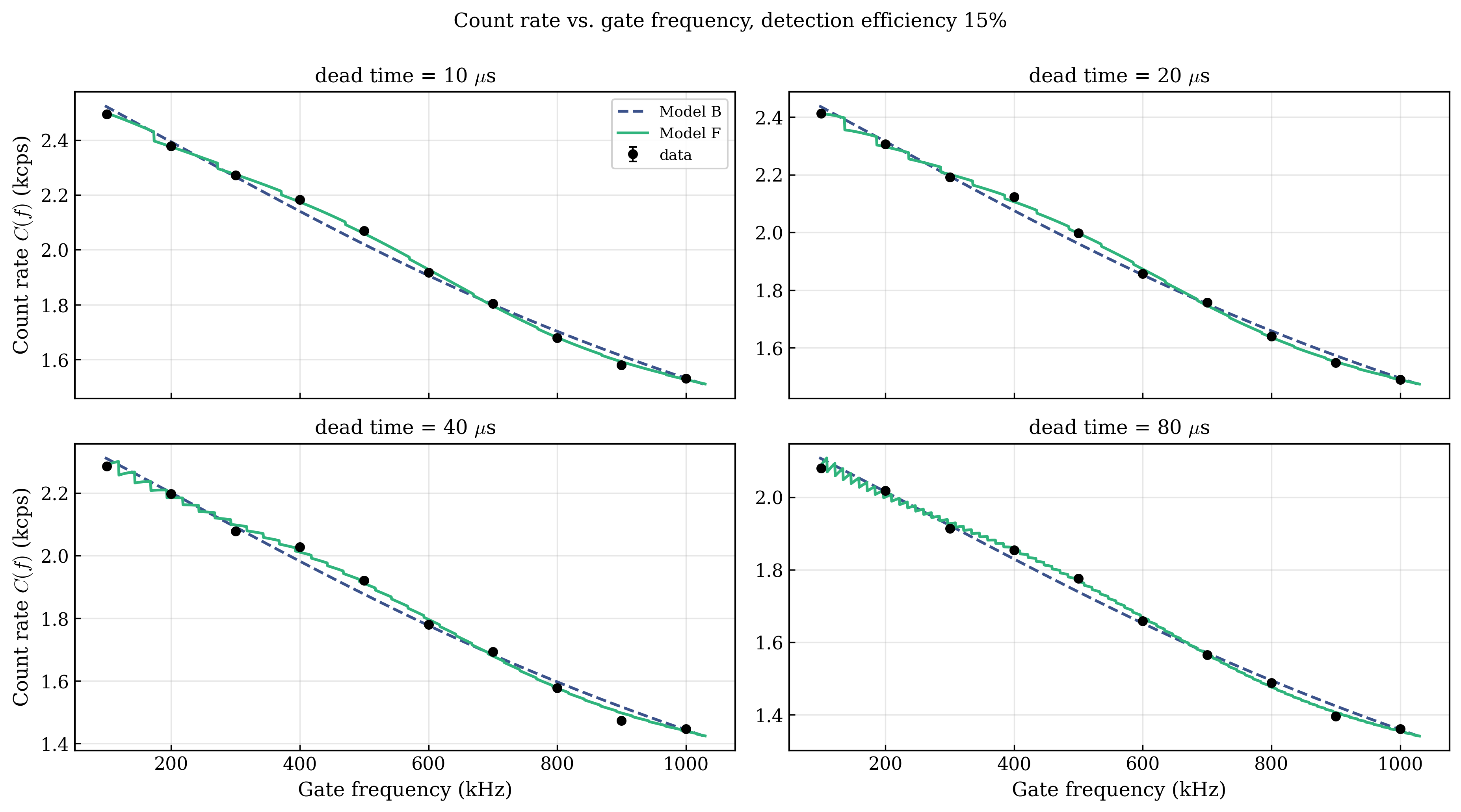}
\caption{Representative $C(f)$ curves at $\eta=15\,\%$ for all four dead times, with Baseline and Full model fits overlaid. The legend in the top-left panel applies to all panels.}
\label{fig:curves}
\end{figure}

Superimposed on the smooth decay of $C(f)$ is a periodic modulation that the Baseline Model cannot reproduce (Fig.~\ref{fig:curves}). Its amplitude, characteristic frequency, and phase, extracted from the Full Model, are collected in Table~\ref{tab:ripple}.

The ripple amplitude decreases significantly between $\eta = 10\,\%$ and $\eta = 20\,\%$ ($8.1\sigma$; Table~\ref{tab:ripple}), consistent with the direction expected if the modulation originates in capacitive feedthrough of the gate waveform into the readout node, coupled through the excess-bias-dependent junction capacitance $C_j(V_{\mathrm{ex}})$ (Sec.~\ref{sec:ripple}): $C_j$ falls as $V_{\mathrm{ex}}$ rises, weakening the coupling. At $\eta = 25\,\%$, however, the amplitude rises again, differing from the $20\,\%$ value by $2.7\sigma$ (Table~\ref{tab:ripple}). A monotonic $C_j(V_{\mathrm{ex}})$ dependence alone therefore does not account for the behavior at the highest bias tested, and a second coupling pathway, or a change of regime, appears to set in there. We present this interpretation as consistent with an electronic and instrumental origin rather than as a confirmed mechanism: it is inferred from the bias dependence of the fitted amplitude, not from a direct measurement of the coupling. A reflection measurement at the gate port as a function of $V_{\mathrm{ex}}$ would test it directly \cite{Losev2022PJ} and is the most economical way to falsify it.

\begin{table*}[!t]
\centering
\caption{Modulation parameters from the Full Model, with the dominant peak of the Lomb-Scargle periodogram of the Baseline residuals shown for comparison. The last column is the relative deviation between the fitted $f_0$ and the
periodogram peak, $|f_0 -f_{\mathrm{peak}}|/f_{\mathrm{peak}}$; the two
quantities are obtained from analyses sharing no parameters.}
\label{tab:ripple}
\begin{tabular}{c c c c c c}
\toprule
$\eta$ (\%) & $a$ (\%) & $f_0$ (kHz) & $\varphi$ (rad) & periodogram peak (kHz) & deviation (\%) \\
\midrule
10 & $3.59 \pm 0.17$ & $863.5 \pm 38.0$ & $-2.06$ & 932.6 & 8.0 \% \\
15 & $1.64 \pm 0.16$ & $718.4 \pm 32.9$ & $-2.69$ & 768.7 & 7.0 \% \\
20 & $1.56 \pm 0.18$ & $704.2 \pm 43.9$ & $-2.56$ & 713.0 & 1.2 \% \\
25 & $2.31 \pm 0.20$ & $723.5 \pm 39.5$ & $-2.36$ & 718.0 & 0.8 \% \\
\bottomrule
\end{tabular}
\end{table*}

The characteristic frequency behaves differently. It drops from $863.5 \pm 38.0$~kHz at $\eta = 10\,\%$ to $718.4 \pm 32.9$~kHz at $\eta = 15\,\%$, a $2.9\sigma$ shift, and then remains constant within uncertainty over the interval $704$ to $723$~kHz for $\eta \geq 15\,\%$. The phase shows no significant trend at all, averaging $-2.42 \pm 0.27$~rad across the four efficiencies. A modulation whose phase is fixed while its amplitude varies by more than a factor of two is what one expects from a single coupling pathway of varying strength, rather than from four unrelated artifacts.

\begin{figure}[htbp]
\centering
\includegraphics[width=\columnwidth]{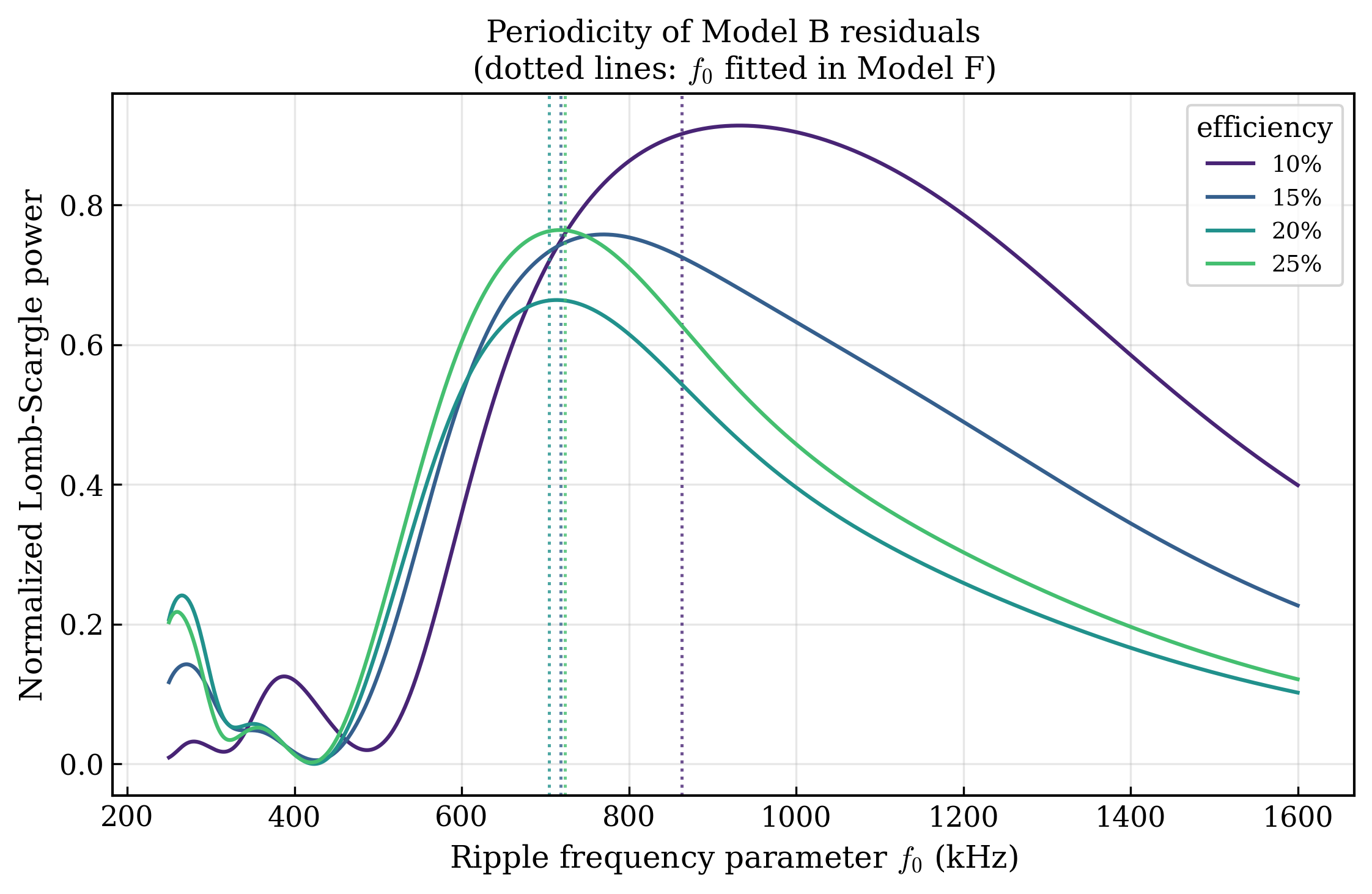}
\caption{Lomb-Scargle periodogram of the Baseline Model residuals, showing peaks near the independently fitted $f_0$ at each efficiency.}
\label{fig:periodogram}
\end{figure}

That the modulation is a property of the measurement and not of the fitting model is established independently by the Lomb-Scargle periodogram \cite{VanderPlas2018} of the Baseline residuals (Fig.~\ref{fig:periodogram}), which contains no sinusoidal component by construction. Its dominant peak agrees with the independently fitted $f_0$ to within $8.0\,\%$, $7.0\,\%$, $1.2\,\%$, and $0.8\,\%$ at $\eta = 10\,\%$ through $25\,\%$, the agreement tightening as efficiency increases. Two analyses that share no parameters converge on the same periodicity.

Taken together with Sec.~\ref{sec:tau_results}, this means that two distinct observables, the recovery time constant and the feedthrough amplitude, arising from separate physical mechanisms, both track the detector operating point. They do so with different functional forms, one linear and monotonic, the other not, so we do not claim a single controlling parameter; but both are consistent with junction properties that evolve continuously with excess bias.


\subsection{Validation of the count-rate model}
\label{sec:validation}

\begin{figure}[htbp]
\centering
\includegraphics[width=\columnwidth]{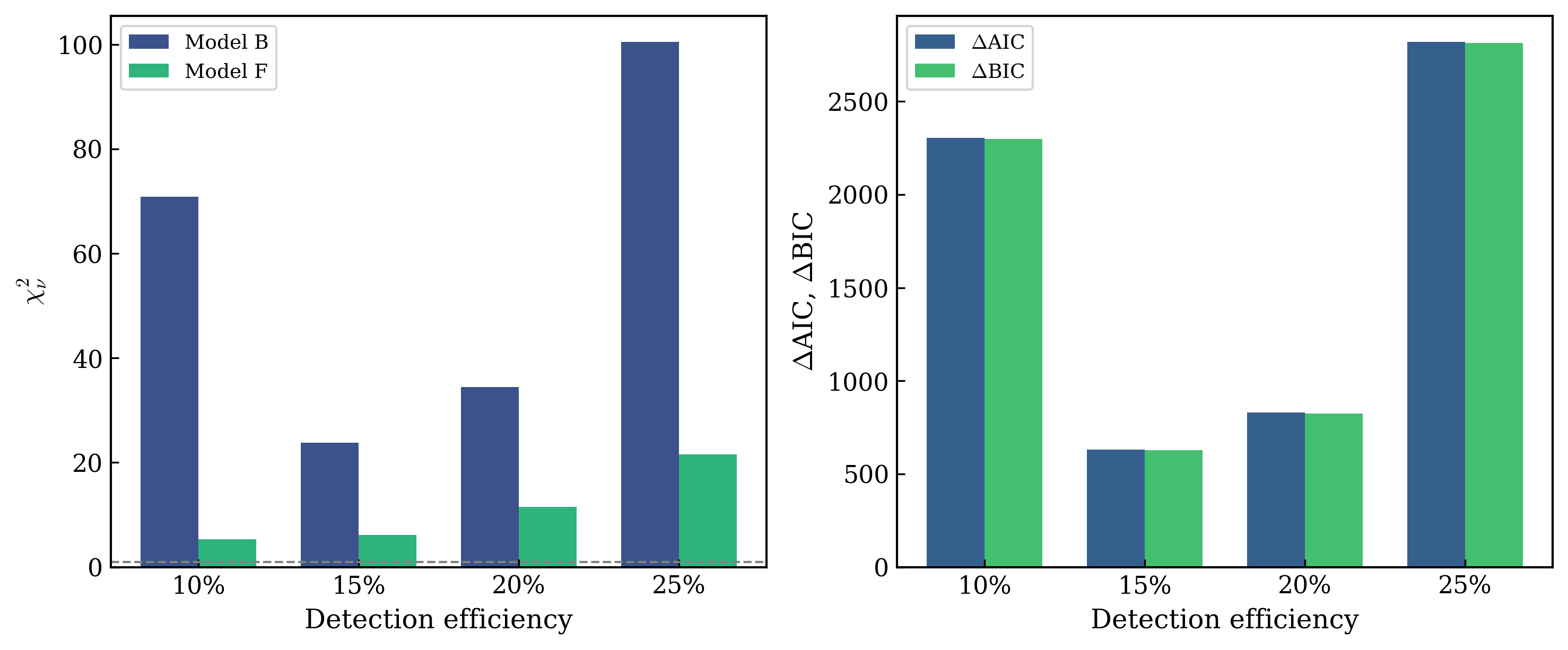}
\caption{$\chi^2_\nu$, $\Delta\mathrm{AIC}$, and $\Delta\mathrm{BIC}$ across the four detection efficiencies.}
\label{fig:metrics}
\end{figure}

Both models reproduce the overall shape of $C(f)$; only the Full Model accounts for the periodic modulation. The information criteria separate them decisively (Fig.~\ref{fig:metrics}, Table~\ref{tab:global_fits}): $\Delta\mathrm{AIC}$ and $\Delta\mathrm{BIC}$ exceed the conventional decisive threshold of $10$ \cite{BurnhamAnderson2004} by two to three orders of magnitude at every efficiency, even after both criteria penalize the Full Model for its three additional parameters. The reduced chi-square falls correspondingly by factors of $13.3$, $3.9$, $3.0$, and $4.6$ (Table~\ref{tab:global_fits}), indicating that the Full Model captures real structure rather than merely absorbing scatter with additional freedom.

These comparisons rest on $\chi^2_\nu$ rather than on $R^2$, which is uninformative here: the Baseline Model already reaches $R^2 > 0.98$ at every efficiency because the smooth dead-time decay dominates the shape of $C(f)$, while the modulation is a small fractional perturbation upon it. With per-point uncertainties reduced by averaging over $N_t$ repetitions, $\chi^2_\nu$ resolves exactly this kind of small systematic residual.

\begin{figure}[htbp]
\centering
\includegraphics[width=\columnwidth]{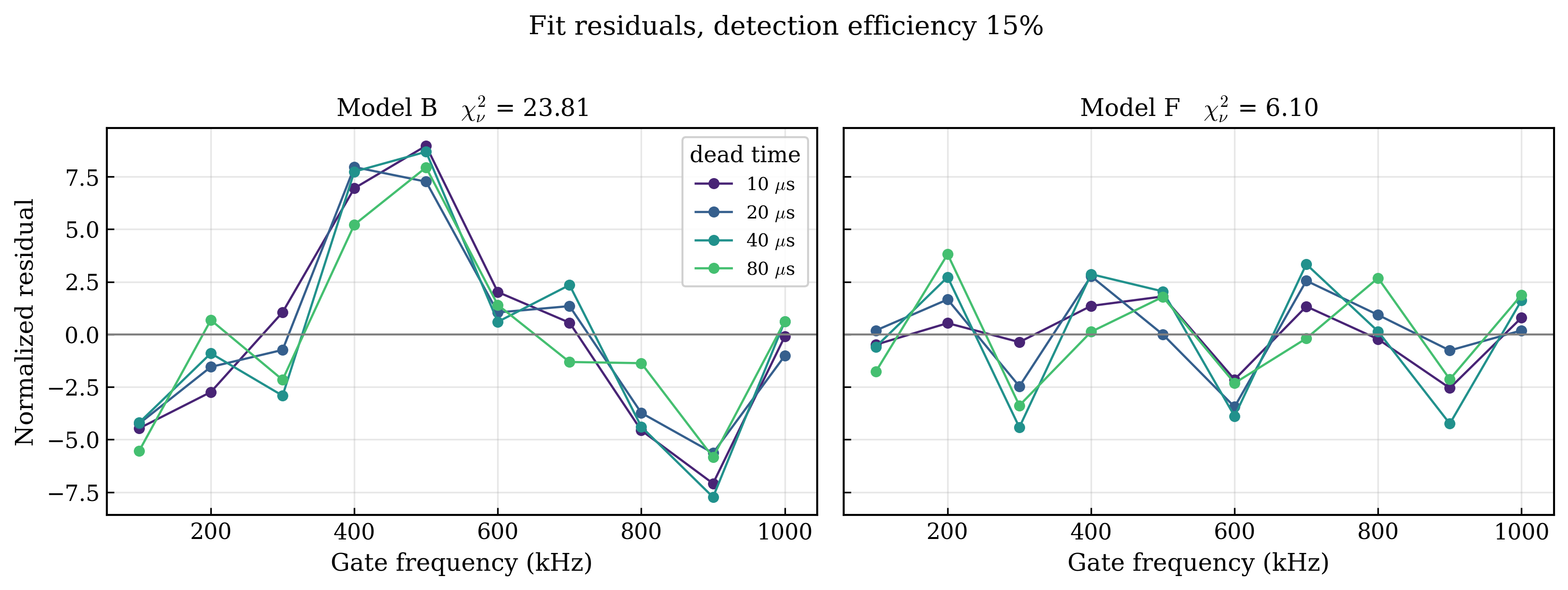}
\caption{Normalized residuals versus gate frequency for the Baseline and Full models at $\eta=15\,\%$.}
\label{fig:residuals}
\end{figure}

The Full Model nonetheless leaves $\chi^2_\nu$ between $5.3$ and $21.6$, and the residual excess grows with efficiency. Normalized residuals (Fig.~\ref{fig:residuals}) show that the periodic structure prominent in the Baseline fit is largely, though not entirely, absorbed. Two contributions plausibly account for what remains. The first is the error model itself: if the $N_t$ sub-measurements per point are not fully statistically independent (Sec.~\ref{sec:methods}), the propagated $\sigma$ understates the true noise and part of the apparent excess reflects an optimistic error bar rather than missing physics. The second is the phenomenological character of the modulation term. A single sinusoid of fixed frequency and phase is the minimal form consistent with the observed periodicity, but the non-monotonic amplitude at $\eta = 25\,\%$ suggests that the true feedthrough response is not fully captured by one harmonic across the whole bias range; a second harmonic, or a frequency-dependent amplitude, are natural extensions to test on a denser grid. We therefore regard the Full Model as the best available description of this detector over this frequency range, and a clear improvement on the dead-time-only treatment, without presenting it as a closed theory. Measured count fluctuations were separately verified to be consistent with the sub-Poisson statistics expected from dead-time regularization (Supplementary Material).

\subsection{Quantized dead time on a commensurate grid}
\label{sec:losttime_results}

The Full Model differs from the Baseline in two respects: the modulation term and the quantization of the effective dead time by the gate period, Eq.~(\ref{eq:taueff}). Evaluating $\tau_{\mathrm{dt}}/T$ at every measurement condition shows that this ratio is an integer at all 160 points of the present grid, a consequence of combining round-number gate frequencies with dead times that are themselves round multiples of $10\,\mu$s. Equation~(\ref{eq:taueff}) therefore reduces identically to $\tau_{\mathrm{eff}} = \tau_{\mathrm{dt}}$ throughout, independently of $\tau_{\mathrm{rec}}$.

This makes the dataset a clean control experiment on the question of what drives the model comparison. Since the lost-time correction is numerically inert at every point, the improvements reported in Sec.~\ref{sec:validation} are attributable entirely to the modulation term, with no possibility of the two mechanisms compensating one another. The corollary is that the quantized dead time itself remains untested by these data, which motivates the deliberately non-commensurate grid proposed in Sec.~\ref{sec:conclusion}.

\subsection{Scope of the present measurements}
\label{sec:scope}

Three boundaries merit explicit mention. First, the excess bias was not measured; we report $\tau_{\mathrm{rec}}(\eta)$ rather than $\tau_{\mathrm{rec}}(V_{\mathrm{ex}})$, so the $C_j \propto V_{\mathrm{ex}}^{-1/2}$ scaling invoked in Sec.~\ref{sec:ripple_results} cannot be tested quantitatively, only in sign. Second, the fully commensurate measurement grid leaves the discrete re-arming term, Eq.~(\ref{eq:taueff}), unexercised. Third, we measure the sum $\tau_{\mathrm{rec}}$, not its constituent blind time and recharge constant, which remain degenerate over the frequency range accessible here (Sec.~\ref{sec:full}); this is a limitation of the count-rate observable rather than of the fit. Separating the two would require gate frequencies comparable to $\tau_{\mathrm{rec}}$ or direct time-stamping of detections, both proposed in Sec.~\ref{sec:conclusion}. Dead times below $10\,\mu$s were excluded, since afterpulsing in this device class remains appreciable at shorter hold-off intervals \cite{Itzler2012,Losev2022JQE}.

Having established $\tau_{\mathrm{rec}}$ as an intrinsic, efficiency-dependent property of the gated detector, it is instructive to compare it with the free-running recovery constant $\tau_r = 112.5$~ns reported by Krause and Walenta \cite{Krause2025} for the same commercial detector model (ID Quantique IDQube NIR). Their fit returned an asymptotic efficiency of $\eta_0 = 19.1$ \%, close enough to our $\eta = 20\,\%$ setting that, given the identical hardware (the
identical part number, IDQube-NIR-FR-MMF-LN, is explicitly reported in their Methods section, confirming rather than merely motivating the comparison), we take it as the fairest available comparison point; the small offset is consistent with the usual gap between a fitted asymptotic efficiency and the corresponding nominal dial-in value, rather than a different operating point. At matched efficiency, our gated $\tau_{\mathrm{rec}} = 202.5 \pm 2.6$~ns exceeds $\tau_r$ by a factor of $1.8$. The same comparison at our other three efficiencies, where no matched free-running value exists, shows the ratio falling from $2.7$ at $\eta = 10$ \% to $2.2$ at $15$ \% and $1.4$ at $25$ \%.

\section{Conclusion}
\label{sec:conclusion}

We have characterized the recovery dynamics of a gated InGaAs/InP single-photon avalanche detector by sweeping the gate frequency and fitting the resulting count-rate curves globally across four dead-time settings. The measurement requires only gate-frequency control and count-rate readout, without a pulsed source, a time-to-digital converter, or any timing hardware beyond the gate generator already present in a gated receiver.

Across detection efficiencies from $10\,\%$ to $25\,\%$, the gated recovery time constant falls linearly from $301$ to $161$~ns, at a rate of $9.28 \pm 0.28$~ns per percentage point, tying the detector operating point quantitatively to the speed at which it re-arms within a gate. The same data reveal a reproducible periodic modulation of the count rate, whose reality is corroborated by an independent periodogram analysis of the fit residuals and whose amplitude tracks the excess bias, as expected of a capacitive coupling at the gate port.

The practical consequence is that an operating point can be chosen with the trade-off between detection efficiency and recovery-limited saturation quantified rather than assumed, in the mode that deployed quantum key distribution receivers actually use. Two further measurements would sharpen the picture: a deliberately non-commensurate frequency and dead-time grid, which requires no new hardware and would exercise the discrete re-arming term left inert here (Sec.~\ref{sec:losttime_results}), and direct time-stamping of detections against the gate edge, which would separate the blind time from the recharge constant that the count rate alone reports only as a sum (Sec.~\ref{sec:full}). Finally, the frequency-sweep method demonstrated here extracts efficiency-resolved recovery dynamics using only standard gate-control and counting electronics, without dedicated single-photon timing hardware.

\section*{Acknowledgment}
The authors declare no conflicts of interest. Claude (Anthropic) assisted with editorial condensation of the Methods and Results sections. All AI-assisted text was reviewed and edited by the authors, who take full responsibility for the content of this article. Supplementary material accompanying this article, including the derivation of the mean click time and quantized effective dead time, is provided with the submission.

\section*{Data Availability}
Data underlying the results presented in this paper are not publicly available at this time but may be obtained from the authors upon reasonable request.

\bibliographystyle{IEEEtran}
\bibliography{bibliography}

\end{document}